\RequirePackage{fixltx2e}
\documentclass[prl,a4paper,twocolumn,superscriptaddress,showkeys]{revtex4}
\usepackage[utf8]{inputenc}
\usepackage[english]{babel}
\usepackage[T1]{fontenc}
\usepackage{amsmath}
\usepackage{amsfonts}
\usepackage{amssymb}
\usepackage{graphicx}

\usepackage{siunitx}
\usepackage{hyperref}

\begin{document}
\author{Hassan Srour}
\affiliation{Laboratoire de Chimie UMR CNRS 5182,
Ecole Normale Supérieure de Lyon/ Université Claude Bernard Lyon1/ Université de Lyon, 
46 Allée d’Italie, 69007 Lyon, France}
\author{Olivier Ratel}
\affiliation{Ingénierie des Matériaux Polymères UMR CNRS 5223 
Universite Jean Monnet de Saint-Etienne/ Université de Lyon; 42023, Saint-Etienne, France}
\author{Mathieu Leocmach}
\affiliation{Laboratoire de Physique UMR CNRS 5672,
Ecole Normale Supérieure de Lyon/ Université Claude Bernard Lyon1/ Université de Lyon,
46 Allée d’Italie, 69007 Lyon, France}
\author{Emma A. Adams}
\author{Sandrine Denis-Quanquin}
\author{Vinukrishnan Appukuttan}
\affiliation{Laboratoire de Chimie UMR CNRS 5182,
Ecole Normale Supérieure de Lyon/ Université Claude Bernard Lyon1/ Université de Lyon, 
46 Allée d’Italie, 69007 Lyon, France}
\author{Nicolas Taberlet}
\author{Sébastien Manneville}
\affiliation{Laboratoire de Physique UMR CNRS 5672,
Ecole Normale Supérieure de Lyon/ Université Claude Bernard Lyon1/ Université de Lyon,
46 Allée d’Italie, 69007 Lyon, France}
\author{Jean-Charles Majesté}
\author{Christian Carrot}
\affiliation{Ingénierie des Matériaux Polymères UMR CNRS 5223 
Universite Jean Monnet de Saint-Etienne/ Université de Lyon; 42023, Saint-Etienne, France}
\author{Chantal Andraud}
\author{Cyrille Monnereau}
\affiliation{Laboratoire de Chimie UMR CNRS 5182,
Ecole Normale Supérieure de Lyon/ Université Claude Bernard Lyon1/ Université de Lyon, 
46 Allée d’Italie, 69007 Lyon, France}
\email{cyrille.monnereau@ens-lyon.fr}

\title{Mediating gel formation from structurally controlled poly(electrolytes) through multiple “head-to-body” electrostatic interactions}

\keywords{Polymers; ionic liquid; gel; self-assembly; rheology}

\begin{abstract}
Tuning the chain-end functionality of a short-chain cationic homopolymer, owing to the nature of the initiator used in the ATRP polymerisation step, can be used to mediate the formation of a gel of this poly(electrolyte) in water. While a neutral end group gives a solution of low viscosity, a highly homogeneous gel is obtained with a phosphonate anionic moiety, as characterized by rheometry and diffusion NMR. This novel type of supramolecular control over poly(electrolytic) gel formation could find potential use in a variety of applications in the field of electroactive materials.
\end{abstract}

\maketitle

Ionic liquids have been used for decades as solvents for a variety of synthetic applications, and have recently experienced a renewed interest in material chemistry~\cite{Antonietti2004,Hapiot2008,Ichikawa2011,MacFarlane2014}. Poly(ionic liquids) (PILs), in particular, show promising characteristics as electro-active materials, more specifically in the field of energy storage application (batteries, dye-sensitized solar cells or fuel cells) \cite{Ohno2001,Kim2011b,Mecerreyes2011,Zhao2011,Yuan2013}. The synthesis of structurally-controlled polymer architectures that incorporate Ionic-Liquid (IL) moieties, and the study of the electrochemical properties of such polymers have become very active topics, as witnessed by the impressive number of recent reports dealing with these two aspects~\cite{Hirao2000,Yuan2011,Matyjaszewski1999,Appukuttan2012}.
The incorporation of PILs into functional devices commonly requires their shaping into processable materials, such as thin films, membranes, or gels. The latter constitute an increasingly sought-after category of materials, as gel poly(electrolytes) are expected to substitute currently used volatile, flammable and leakage-prone organic solvent-based systems in future energy storage devices~\cite{Wang2003,Choudhury2009,Lewandowski2004}.

Different strategies have been proposed to mediate gel formation from a variety of polyelectrolytes~\cite{Katsampas2005,Xiong2012,Hu2014}. In most cases these strategies rely on the physical interaction of colloidal nanobjects (nanoparticles, micelles, highly crosslinked polymer networks) resulting in scattering colloidal gels with relatively poor mechanical and diffusion properties. By comparison, gels based on the supramolecular interactions of short polymeric chains constitute a very promising class of materials for practical applications, as they combine the advantages of polymeric materials (processability, solid-like behavior) and individual molecule solutions (efficient diffusion of solvent molecules, dynamic reorganization, and self-healing properties)\cite{Noro2011,Li2014,Furusho2014}. So far the few reported examples dealing with the formation of such ``supramolecular'' gels from polycations (anions) have all relied on a similar strategy, i.e. the addition of a molecule or a short polymer bearing multiple anionic (cationic) groups, which induces gel formation by the occurrence of a non-covalent crosslinking network~\cite{Zhao2013,Becht2011}.

In this paper, we propose a markedly different and more straightforward conceptual approach for the making of supramolecular homopolymer gels, based on electrostatic interactions between the positive charges held by the repeating cationic units (imidazolium) and a terminal anionic group (phosphonate) introduced in the initiation step of the atom transfer radical polymerisation (ATRP). We bring conclusive evidence that the single, relatively weak interaction between both entities is sufficient to mediate the formation of a gel in water, while in the absence of this interaction the aqueous polymer solution behaves like a Newtonian fluid of low viscosity.

\begin{figure}
\includegraphics[]{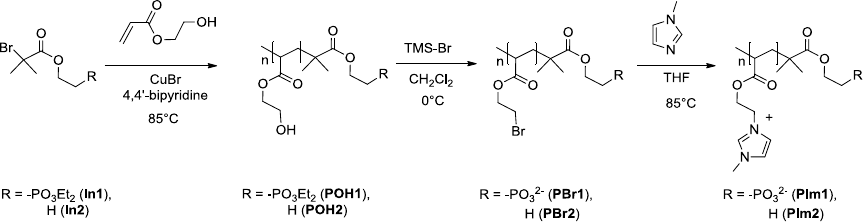}
\caption{Syntheses of PIm1 and PIm2 and of their intermediate.}
\label{fig:reaction}
\end{figure}

Tunability of the polymer chain end, which is a keystone element in the obtention of those gels, is a direct benefit of the methodology we specifically developed for the synthesis of our PILs: as an intrinsic feature of the ATRP used in the first step of our synthesis relies it is indeed possible to control the chemical nature of the polymer chain-end thanks to the molecular structure of the so-called polymerization initiator making it possible to readily incorporate any functionality. Thus, we chose to work with two different bromoisobutyryl ester derivatives: the phosphonate terminated one (In1) is used to provide the anion/cation interaction that is required to mediate gel formation, while the ethyl terminated one (In2) serves as a blank. In the case of In1, the phosphonate moiety was introduced in its ethyl-protected form, using a straightforward two-step methodology. Then, the ethylphosphonate- (PIm1) and ethyl-terminated (PIm2) polymers were synthesized according to the same general methodology as described in a previous report\cite{Appukuttan2012}. This methodology relies on the following three-step sequence consisting of i) ATRP of hydroxyethylacrylate followed by ii) nucleophilic substitution of the pendant hydroxyl group with bromide ions and iii) addition of methylimidazole, resulting in quaternarization of the nucleophile (Figure~\ref{fig:reaction}). Polymerization times were adjusted to get similar polymer chain lengths in both cases (n \textit{ca} 70 for POH1 and POH2), which was confirmed by \textsuperscript{1}H NMR and Steric exclusion chromatography (SEC) experiments. Deprotection of the phosphonic ester in POH1 could be readily and quantitatively achieved at the same step as the nucleophile OH-to-Br substitution, making the whole procedure leading to PBr1 extremely straightforward. POH2 was converted to PBr2 using this same procedure. After introduction of the imidazolium group, PIm1 and PIm2 could be isolated as white rubbery materials. Experimental details are provided as SI (S3-7) 

\begin{figure}
\includegraphics[]{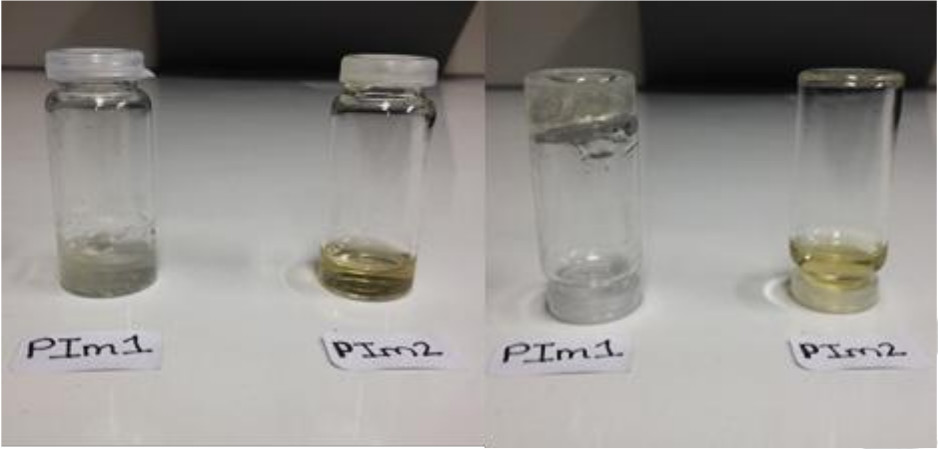}
\caption{Pictures of 10\% weight solutions of PIm1 and PIm2 in water before (left) and after (right) flipping of the vial.}
\label{fig:flip}
\end{figure}

When mixed with water at a weight concentration between typically 5\% to 30\%, it was noticed that PIm1 forms a very homogeneous and optically transparent (non-scattering) gel which does not flow under its own weight whereas PIm2 seems to behave like pure water (Figure~\ref{fig:flip}).

\begin{figure}
\includegraphics{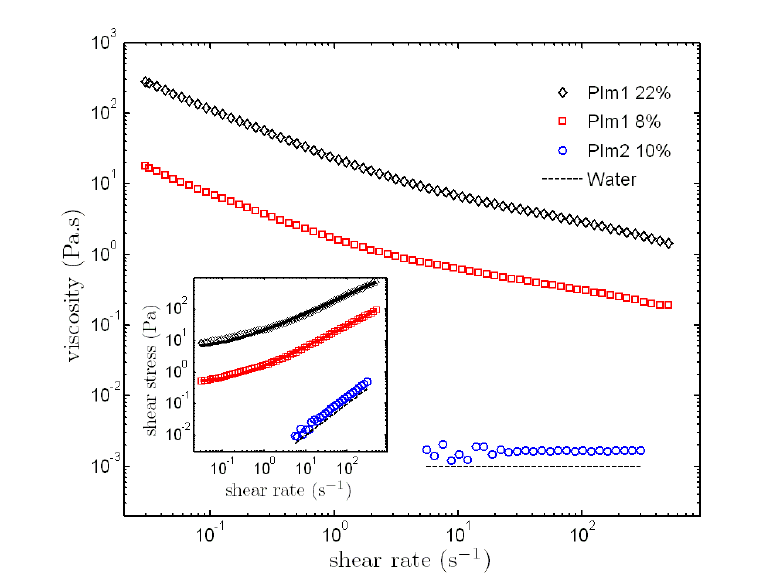}
\caption{Shear rate dependence of the viscosity for PIm1 at 22\% (black) and 8\% (red) weight concentration and PIm2 at 10\% (blue). The black dashed line is the viscosity of pure water. Inset: Shear rate dependence of the shear stress. Continuous lines are best fits by a Herschel-Buckley model with a concentration-dependent yield stress. Note that the measurement range for PIm2 is smaller than that of PIm1 due to the resolution limit of the instrument (for stresses below \SI{1e-2}{\pascal}).}
\label{fig:yieldstress}
\end{figure}

In order to characterize this behavior standard rheometry experiments were conducted. Figure~\ref{fig:yieldstress} shows the viscosity $\eta$ at different applied shear rates  in a cone-and-plate geometry at a fixed temperature of \SI{25}{\celsius}\cite{Macosko1994}. The difference between PIm1 and PIm2 is obvious. The viscosity of PIm2 is constant and close to that of pure water at \SI{25}{\celsius} showing that PIm2 behaves as a low-viscosity Newtonian fluid. Conversely, the viscosity of PIm1 always remains several orders of magnitude higher than that of PIm2, increases with decreasing shear rate (a behavior referred to as ``shear-thinning'') and keeps increasing without showing any sign of a Newtonian plateau at low shear rates. This divergence of the viscosity at vanishingly small shear rates indicates a solid-like behavior of the sample at rest as typically found in gels. More precisely, the inset in figure~\ref{fig:yieldstress} shows the shear stress $\sigma=\eta\dot{\gamma}$ as a function of the applied shear rate. Contrary to the case of PIm2, the shear stress for PIm1 tends towards a finite value $\sigma_c$ at low shear rates. This indicates that a critical force ($\sigma_c$) is required for the material to flow. This critical value is referred to as the ``yield stress'' in rheology\cite{Barnes1999} and ascertains the gel-like behavior of PIm1. The solid lines in the inset are the best fits to the data using a Herschel-Buckley law $\sigma = \sigma_c + A\dot{\gamma}^n$ and allow one to measure the values of the yield stress. It was found that PIm1 is a weak gel, with $\sigma_c = \SI{4.7}{\pascal}$ for a weight concentration of 22\% and $\sigma_c = \SI{0.4}{\pascal}$ for 8\%. 

\begin{figure}
\includegraphics[]{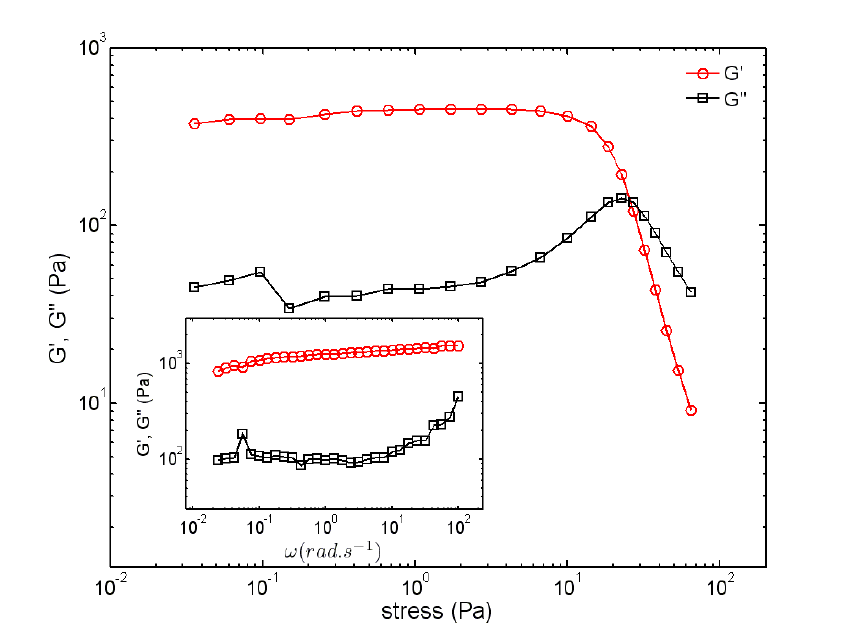}
\caption{Storage ($G^\prime$) and loss ($G^{\prime\prime}$) moduli during oscillatory shear experiments for  PIm1 at 20 \%wt. $G^\prime$ (red) and $G^{\prime\prime}$ (black) vs stress amplitude during an amplitude sweep at a fixed angular frequency $\omega=\SI{0.1}{\radian\per\second}$. Inset: $G^\prime$ (red) and $G^{\prime\prime}$ (black) vs angular frequency $\omega$ during a frequency sweep at a fixed strain amplitude of 0.1\% (corresponding to a stress amplitude of \SI{1}{\pascal}).}
\label{fig:linearrheol}
\end{figure}

The gel-like nature of PIm1 deduced from steady shear was also confirmed through oscillatory shear rheology using both amplitude sweeps at constant frequency and frequency sweeps within the linear regime. At a fixed angular frequency of \SI{0.1}{\radian\per\second}, PIm1 at 20 \% wt. shows an elasticity-dominated behavior ($G^\prime\gg G^{\prime\prime}$) until a stress amplitude of nearly \SI{20}{\pascal} and a liquid viscosity-dominated behavior ($G^\prime\ll G^{\prime\prime}$) above this value (Figure~\ref{fig:linearrheol}). At a stress amplitude of \SI{1}{\pascal}, a frequency sweep on PIm1 at 20 \% shows a behavior typical of a polymer network or soft glassy system\cite{Larson1999}, with nearly constant storage and loss moduli ($G^\prime\approx \SI{1000}{\pascal}$ and $G^{\prime\prime} \approx \SI{100}{\pascal}$ respectively, inset of Figure~\ref{fig:linearrheol}). Note that a shear stress of \SI{1}{\pascal} is well below $\sigma_c$, consistent with the solid-like behavior shown in Figure~\ref{fig:yieldstress}. A discrepancy still exists between the critical shear stress observed in steady shear and that observed in amplitude sweeps. This could be partly due to the definition chosen for the yield stress and to the different mode of solicitation\cite{Coussot2014}. In particular the loss modulus $G^{\prime\prime}$ is seen to increase significantly for stresses amplitudes above \SI{5+-1}{\pascal}, which is closer to the yield stress estimated above. In other systems such an increase followed by a maximum of the loss modulus has been attributed to structural rearrangements before the system yields\cite{Hyun2011}. In any case Figure~\ref{fig:linearrheol} clearly demonstrates the existence of a weak gel whose non-covalent crosslinks are greatly strain/stress-sensitive as similarly observed for filled polymers\cite{Harwood1965}. This property can be advantageously used, as it allows processing of the material through simple injection (video 1).

\begin{figure}
\includegraphics[]{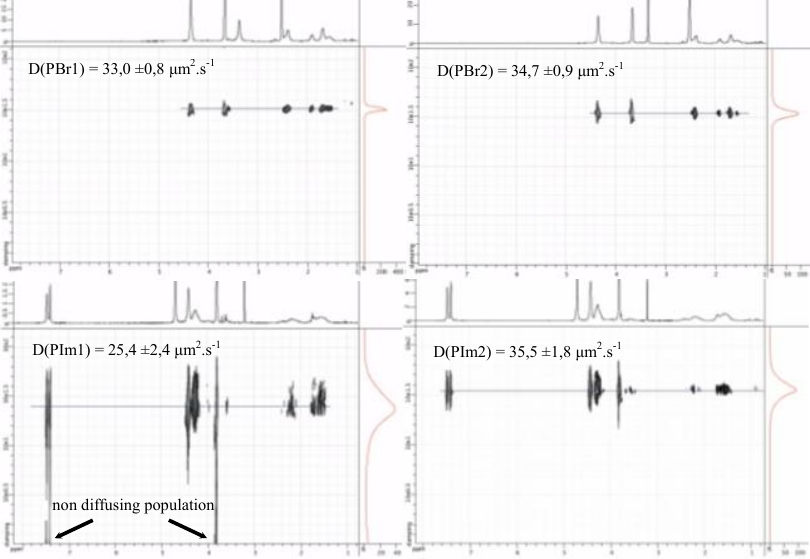}
\caption{DOSY spectra of \SI{10}{\gram\per\litre} solutions of PBr1 and PBr2 in d6-DMSO (top), PIm1 and PIm2 in D\textsubscript{2}O.}
\label{fig:DOSY}
\end{figure}

This striking difference in the behavior of PIm1 and PIm2 upon solubilization in water could be further demonstrated and characterized using 2D diffusion NMR (DOSY) (Figure~\ref{fig:DOSY}).

Both hydroxyl (POH1 and POH2, S9) and brominated precursors (PBr1 and PBr2) show diffusion coefficients that are, within error margin, undistinguishable from one another. From the diffusion data, it is possible to extract polydispersity indexes (PDIs)\cite{Vieville2011}. PDIs of 1.24 and 1.25 are found for POH1 and POH2 in D\textsubscript{2}O, respectively; this is slightly higher than what is given by SEC analyses in DMSO (1.14 and 1.16, resp.; S8). Similarly, values of 1.15 and 1.14 are measured for PBr1 and PBr2 in d6-DMSO. 

In contrast, DOSY spectra of PIm1 and PIm2 show marked differences. First the average diffusion coefficient of the diffusing polymer chain species of PIm1 is decreased by a 1.4 ratio (from \SIrange{35}{25}{\square\metre\per\second}) which clearly exceeds the error margin of the experiment. A possible explanation would be the significant difference in the weight of the two polymers. For a linear polymer chain of molecular weight $M$, the diffusion coefficient $D$ is assumed to obey the relation $M\propto D^{-d_F}$ with $d_F$, the fractal dimension of the polymer chain ($d_F$ values between $5/3$ for stretched-out, $3$ for collapsed polymer chains).  Considering a typical value $d_F = 2$, this result would suggest an $M$ value about twice as high for PIm1 as for PIm2.

Since PIm1 and PIm2 have similar molecular weights and compositions, attested by multiple SEC and NMR analyses of their synthesis precursors, the difference between their diffusion coefficients shows that PIm1 does not behave like a single, linear non-interacting polymer. For this sample the measured diffusion coefficient corresponds to a weighted average between the free and interacting chains, the exchange rate being fast compared to the diffusing time (\SI{200}{\milli\second}). Furthermore the presence of non-diffusing signals suggests that a significant proportion of the polymer chains are already immobilized, although DOSY experiments were recorded at polymer weight concentrations in D\textsubscript{2}O that were significantly lower than the critical gel formation concentration.

\begin{figure}
\includegraphics[]{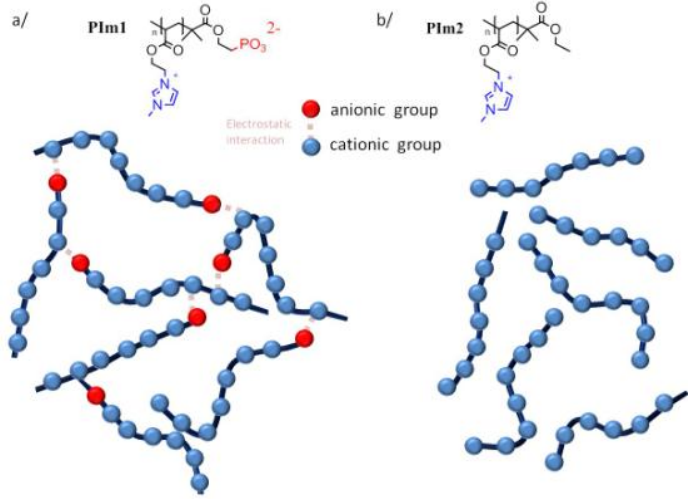}
\caption{Proposed model accounting for the difference in behavior between PIm1 (left) and PIm2 (right); when the polyelectrolyte is terminated by a phosphonate ion, interaction with imidazolium moieties from proximal polymer chains induces the formation of a supramolecular network. In the absence of this interaction, polymer chains are freely flowing within the solution.}
\label{fig:model}
\end{figure}

Altogether, these results suggest that the presence of this single terminal phosphonic acid at the extremity of each short polymer chain is enough to mediate efficient gel formation. This behavior can be rationalized on the basis of the model illustrated in Figure~\ref{fig:model}: owing to a single electrostatic interaction between their terminal phosphonate and lateral imidazolium group, each polymer chain is involved in a global supramolecular network; this gives rise, above a critical concentration limit, to cohesion forces that ``freeze'' the system and prevent it from flowing. 

The presence, along each polymer chain, of multiple cationic anchoring points for the single anionic end allows dynamic reorganization processes within the gel network. This provides i) stabilization of the gel structure in spite of the relatively weak single electrostatic interaction that connects each polymer chain to its neighbor and ii) avoids the formation of micro-domains, which makes the gel optically transparent. The nature of these ionic interactions was further assessed by complementary experiments: it was noticed that variations of pH and/or ionic strength considerably affected the polymer organization; while moderate acidification (HCl, pH=4) preserved the gel structure, basification (NaOH, pH=10), turned it into a viscous liquid. Besides, upon using physiological serum (NaCl aq, 0.9\% wt.), no gel formation was observed and an inhomogeneous solution of low viscosity was obtained (S10). Although further investigations will be required to get more insight on the precise underlying mechanisms, these preliminary experiments strongly suggest that screening of the interacting charges by the excess ions is involved in the gel collapse process. This property might open several perspectives related to the use of similar systems as stimuli responsive materials.

We believe that the concept introduced in this paper constitutes a straightforward alternative for the production of poly(electrolyte) gels, which could find potential use in a broad range of applications related to electrolytic materials and devices. Although illustrated herein in the specific case of electrostatic interaction between complementary anionic and cationic group, we hope that it could be extended to other types of supramolecular interactions, for instance with initiator/monomer combinations involving complementary H-bond donor/acceptor units.

\begin{acknowledgments}
The authors thank the Region Rhône Alpes and the Programme d'avenir Lyon- Saint Etienne (PALSE NoGELPo) for fundings and postoctoral grants for HS and ML. The Region Rhône Alpes (ARC 6-Energies program) is also acknowledged for a doctoral grant (OR). SM and NT acknowledge funding from the European Research Council under the European Union's Seventh Framework Program (FP7/2007-2013) / ERC grant agreement No. 258803. Dr. Marc-André Delsuc is warmly acknowledged for discussions regarding DOSY results.
\end{acknowledgments}

\bibliography{../../bib/PALSE.bib}

\end{document}